\UseRawInputEncoding
\documentclass[
8pt,
superscriptaddress,
prx, reprint,
amsmath,amssymb,
aps, 
]{revtex4-2}

\usepackage{graphicx}
\usepackage{dcolumn}
\usepackage{bm}
\usepackage{siunitx}
\usepackage{tcolorbox}
\usepackage{braket}

\begin{document}
\title{\textbf{Entangling remote qubits through a two-mode squeezed reservoir}}
\author{A. Andr\'es-Juanes}
\email{Contact author: Alejandro.AndresJuanes@ist.ac.at}
\affiliation{
 Institute of Science and Technology Austria (ISTA), Am Campus 1, 3400 Klosterneuburg, Austria}
\author{J. Agust\'i}
\affiliation{Technical University of Munich, TUM School of Natural Sciences, Physics Department, 85748 Garching, Germany}
\affiliation{Walther-Meißner-Institut, Bayerische Akademie der Wissenschaften, 85748 Garching, Germany}
\affiliation{Munich Center for Quantum Science and Technology (MCQST), 80799 Munich, Germany}
\affiliation{Institute of Fundamental Physics IFF-CSIC, Calle Serrano 113b, 28006 Madrid, Spain}
\author{R. Sett}
\affiliation{
 Institute of Science and Technology Austria (ISTA), Am Campus 1, 3400 Klosterneuburg, Austria}
 \author{E. S. Redchenko}
\email{Current address: Vienna Center for Quantum Science and Technology, Atominstitut, TU Wien, 1020 Vienna, Austria}
\affiliation{
 Institute of Science and Technology Austria (ISTA), Am Campus 1, 3400 Klosterneuburg, Austria}
 \author{L. Kapoor}
\affiliation{
 Institute of Science and Technology Austria (ISTA), Am Campus 1, 3400 Klosterneuburg, Austria}
 \author{S. Hawaldar}
\affiliation{
 Institute of Science and Technology Austria (ISTA), Am Campus 1, 3400 Klosterneuburg, Austria}
\author{P. Rabl}
\affiliation{Technical University of Munich, TUM School of Natural Sciences, Physics Department, 85748 Garching, Germany}
\affiliation{Walther-Meißner-Institut, Bayerische Akademie der Wissenschaften, 85748 Garching, Germany}
\affiliation{Munich Center for Quantum Science and Technology (MCQST), 80799 Munich, Germany}
\author{J. M. Fink}
\email{Contact author: jfink@ist.ac.at}
\affiliation{
 Institute of Science and Technology Austria (ISTA), Am Campus 1, 3400 Klosterneuburg, Austria}
\date{\today}

\begin{abstract}
\bf{The distribution of entanglement across distant qubits is a central challenge for the operation of scalable quantum computers and large-scale quantum networks. 
Existing approaches rely on deterministic state transfer schemes or probabilistic protocols that require active control or measurement and postselection. 
Here we demonstrate an alternative, fully autonomous process, where two remote qubits are entangled through their coupling to a quantum-correlated photonic reservoir. In our experiment, a Josephson parametric converter produces a Gaussian, continuous-variable entangled state of propagating microwave fields that drives two spatially separated superconducting transmon qubits into a stationary, discrete-variable entangled state. 
Beyond entanglement distribution, we also show that superconducting qubits can be used to directly certify two-mode squeezing, with higher sensitivity and without the need for calibrated noise-subtraction.
These results establish networks of qubits interfaced with distributed continuous-variable entangled states as a powerful new platform for both foundational studies and quantum-technology relevant applications.}
\end{abstract}

\maketitle
\newpage

\section*{Introduction}
Entanglement is the distinct feature that separates quantum mechanics from classical theories and the key resource that underlies most quantum information processing and quantum communication paradigms~\cite{horodecki_quantum_2009}. Once established across two or multiple locations, entanglement can be further purified~\cite{bennett_purification_1996} and harnessed for secure quantum communication~\cite{
gisin_quantum_2007, northup_quantum_2014}, quantum state teleportation~\cite{bennett_teleporting_1993}, or the remote execution of quantum gates~\cite{gottesman_demonstrating_1999} using classical communication only. The distribution of entanglement is thus of fundamental  importance for the operation of large quantum networks and scalable quantum computing platforms.  

Entanglement comes in various forms, each with different practical utility. While continuous-variable (CV) entangled states~\cite{braunstein_quantum_2005,weedbrook_gaussian_2012, usenko_continuous-variable_2025} of propagating optical or microwave fields can be efficiently generated based on weak 
optical~\cite{ou_realization_1992}, microwave~\cite{eichler_observation_2011}, mechanical~\cite{barzanjeh_stationary_2019} or electro-optic~\cite{sahu_entangling_2023} nonlinearities, with high throughput and distributed over long distances, most applications require the entanglement of stationary qubits, i.e., discrete-variable (DV) systems, for further processing. Surprisingly, it has been found that this apparent mismatch between readily available and practically useful entangled states can be overcome by coupling qubits to a broadband reservoir of correlated photonic states~\cite{kraus_discrete_2004, paternostro_complete_2004, didier_remote_2018,govia_stabilizing_2022,agusti_long-distance_2022}. Through this process, the qubits are driven --- fully autonomously and without direct interaction --- into a pure and almost maximally entangled Bell state. 

Compared to other actively controlled~\cite{humphreys_deterministic_2018,axline_-demand_2018,kurpiers_deterministic_2018,campagne-ibarcq_deterministic_2018,leung_deterministic_2019,zhong_deterministic_2021,almanakly_deterministic_2025}, 
heralded~\cite{narla_robust_2016,dickel_chip--chip_2018,pompili_realization_2021, teoh_robust_2025}, 
measurement- and feedback-based~\cite{roch_observation_2014, liu_comparing_2016}, or autonomous~\cite{krauter_entanglement_2011,shah_stabilizing_2024-1} 
entanglement generation protocols where photons are exchanged between the first and the successive nodes, the distribution and transfer between CV and DV entanglement realized in this work relies on an intriguing non-local interference effect that harnesses the preexisting quantum correlations of the distributed photonic state. Therefore, under ideal conditions, this scheme can be applied over arbitrary distances and extended to complex, multi-qubit entangled states using only a single correlated photon source~\cite{agusti_autonomous_2023}.

In this work, we present the first experimental demonstration of this hybrid  entanglement distribution scheme by driving two separated transmon qubits with the output of a non-degenerate Josephson parametric converter (JPC) \cite{roch_widely_2012, abdo_nondegenerate_2013}. 
The JPC produces a propagating two-mode squeezed (TMS) state of microwave photons \cite{flurin_generating_2012}, which successively relaxes the two frequency-detuned qubits, which are separated from the photon source by 50 cm of coaxial cable each, into an entangled steady state. 

We verify and quantify the predicted transfer of entanglement from a CV reservoir to a DV qubit state together with the underlying non-local interference mechanism. The observed build-up, stabilization, and squeezing-dependent concurrence of the reduced two-qubit state of up to $\mathcal{C}=0.10 \pm 0.01$ is fully consistent with a theoretical model~\cite{agusti_long-distance_2022} and we point out clear pathways for further improvements and extensions to multi-qubit settings.

We further show that by employing the qubits as very efficient correlation detectors, we can certify entanglement of a weakly excited TMS microwave state, directly at cryogenic temperatures and in a parameter regime where conventional linear 
detection schemes with noisy pre-amplifiers 
are inefficient and require calibrated noise subtraction \cite{eichler_observation_2011,flurin_generating_2012,menzel_path_2012,Ku_generating_2015}.

\section*{Entanglement protocol}
\begin{figure}[t]
    \centering
    \includegraphics[width=\columnwidth]{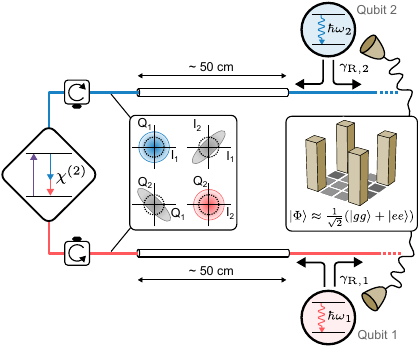}
    \caption{\textbf{Sketch of the implemented entanglement distribution scheme.} A JPC generates a TMS state (left box: Wigner representation of the microwave field. Dashed circle indicates the vacuum level) that propagates in a cascaded configuration to two distant qubits (blue and red shaded circles). Driven by these correlated fields, the qubits relax into the maximally entangled Bell state $\ket{\Phi^+}$ (right box: density matrix elements of the joint qubit state) and are read out dispersively via separate resonators (not shown).
    }
    \label{fig:1}
\end{figure}

We implement a prototype dual-rail quantum network as depicted in Fig.~\ref{fig:1}, where a JPC acts as an entanglement source that emits a broadband TMS state of correlated microwave fields. The two output ports of the JPC are each connected via a coaxial cable to one of the two transmon qubits located on a second chip (see Appendix~\ref{app: setup} for more details about the experimental setup). The qubits decay symmetrically with 
rates $\gamma_{\mathrm{L}, i} = \gamma_{\mathrm{R}, i}$, where $i=1,2$ labels the qubits, into both the left- and right-propagating waveguide modes, but they are separated from the JPC by a circulator to prevent any backaction or any direct qubit-qubit interactions via the transmission lines. 

The JPC consists of two stripline-based resonators with frequencies $\omega_1/(2\pi)=\SI{6.761}{\GHz}$ and $\omega_2/(2\pi)= \SI{10.044}{\GHz}$, which are coupled by a Josephson ring modulator \cite{abdo_nondegenerate_2013} and pumped by an external driving field of frequency $\omega_{\rm p}= \omega_1+\omega_2$. The resulting down-conversion process can be described by an effective two-mode squeezing Hamiltonian $(\hbar=1$) 
\begin{equation}\label{eq:HJPC}
    H_{\rm JPC} = i \frac{\sqrt{\kappa_1 \kappa_2}}{2} \epsilon_{\rm p}\left( e^{i\phi_{\rm p}} a_1^\dagger a_2^\dagger -  e^{-i\phi_{\rm p}} a_1 a_2\right),
\end{equation}
where $a_i$ ($a_i^\dag$) are the bosonic annihilation (creation) operators and  $\kappa_i$ the decay rates of the two modes. The parameters  $\epsilon_{\rm p}\in [0,1)$ and $\phi_{\rm p}$ determine the dimensionless strength and the phase of the pump field, respectively. The JPC produces two Gaussian output fields, which are centered around $\omega_1$ and $\omega_2$ and are correlated over a bandwidth that is set by $(\kappa_1, \kappa_2)/(2\pi) = (60, 75)$ MHz. 
As indicated in the left inset of Fig.~\ref{fig:1}, within this bandwidth, the variance of the joint quadratures of the fields, $\mathrm{Var}\{I_1 + I_2\} = \mathrm{Var}\{Q_1 - Q_2\} < 1/2$, are squeezed below the vacuum level, which, according to the Duan-Simon criterion~\cite{duan_inseparability_2000,simon_peres-horodecki_2000}, certifies entanglement (see Appendix \ref{app: TMS characterization} for a detailed characterization of the JPC output fields).

In our setup, each of the correlated outputs of the JPC drives one of the otherwise decoupled qubits. In the broadband limit $\kappa_1,\kappa_2 \gg \gamma_{\mathrm{R},1}, \gamma_{\mathrm{R},2}$, this scenario implements the coupling of the qubits to a correlated photonic reservoir, which, under otherwise ideal and symmetric conditions, relaxes the qubits into a pure stationary state $\rho(t\rightarrow \infty) = |  \Phi\rangle \langle \Phi |$ with~\cite{kraus_discrete_2004}
\begin{equation}\label{eq:qubit_state}
    \ket{\Phi} = \frac{\sqrt{N + 1}\ket{gg} + e^{i\phi_{\rm p}}\sqrt{N}\ket{ee}}{\sqrt{2N + 1}}.
\end{equation}
This state approaches a maximally entangled Bell state for $N \gtrsim 1$, where $N=\sinh^2(r)$ corresponds to the characteristic photon number for a TMS state with squeezing parameter $r$. In view of the different Hilbert-space dimensions and excitation numbers involved, this autonomous and almost ideal extraction of DV entanglement from a CV photonic state is rather unexpected. Indeed, it relies on a non-local interference effect, which can be understood by adopting the simplified two-mode form, $|\Psi_{\rm TMS}\rangle\sim \sum_{n}[\tanh{(r)}]^n|n\rangle_1 |n\rangle_2$, for the state of the microwave fields in the Fock basis. This state couples to the qubits via a Jaynes-Cummings interaction 
$H_{\rm JC}\sim i (a^\dagger_1 \sigma^-_1-a_1 \sigma^+_1+a^\dagger_2 \sigma^-_2-a_2 \sigma^+_2)$ and it can be readily verified that 
\begin{equation}\label{eq:DarkStateCondition}
H_{\rm JC} \ket{\Phi} |\Psi_{\rm TMS}\rangle=0.
\end{equation} 
Therefore, $|\Phi\rangle$ is the unique entangled dark state that decouples from the correlated photonic environment. Interestingly, the dark-state condition in Eq.~\eqref{eq:DarkStateCondition} arises from the destructive interference, for example, between a photon emission event, $\sim \sigma_1^- a_1^\dag$, in the first waveguide and a photon absorption process, $\sim \sigma_2^+ a_2$, at the location of the second qubit, while both qubits can be arbitrarily far apart. 

\begin{figure*}[t]
    \centering
    \includegraphics{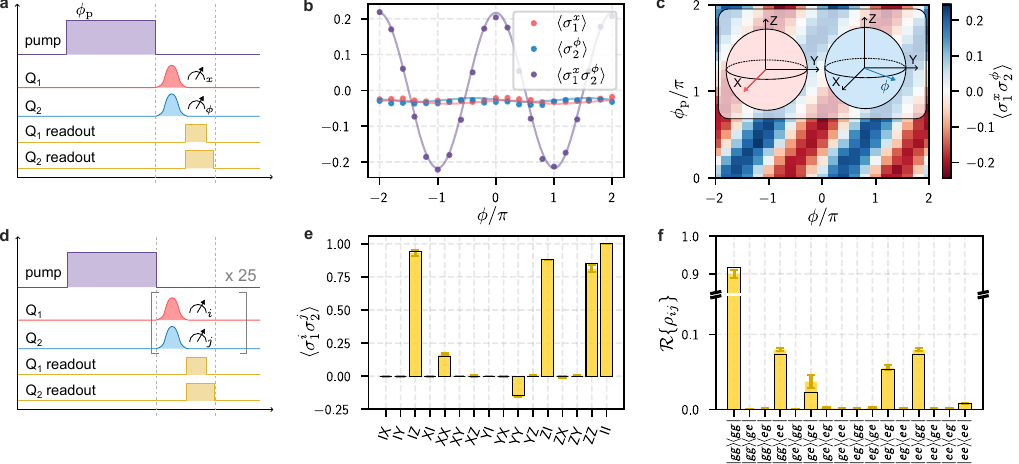}
    \caption{\textbf{Phase coherence and qubit tomography.}
    \textbf{a}, Pulse sequence for the pump phase sweep experiment. Phase-controlled JPC pump pulse (violet), qubit control pulses sent through the resonators (red and blue) controlling the measurement axis $\phi$, and qubit readout pulses (yellow) applied to the readout resonators are shown. \textbf{b}, The measured expectation values of the single and two-qubit operators (see legend) are shown as a function of the qubit 2 measurement axis $\phi$. Lines are fits to a cosine function. 
    \textbf{c}, The same joint expectation value as a function of the phase of the applied JPC pump, $\phi_{\rm p}$. The inset depicts the orientation of the measurement axis. 
    \textbf{d}, Pulse sequence for the qubit tomography experiment implementing 25 different measurement bases for a fixed pump phase.
    \textbf{e}, Averaged measured expectation values $\langle\sigma_1^i \sigma_2^j\rangle$ and 
    \textbf{f}, real part of the reconstructed 2-qubit density matrix $\rho$ for a pump drive of amplitude $\epsilon_{\rm p}^\star = 0.25$ (yellow bars). Error bars show the $1\sigma$ standard error from 7 repetitions with $10^7$ averages each. Solid frames correspond to the theory model (see Appendix~\ref{app: theory}).}
    \label{fig:2}
\end{figure*}

\section*{Results}
To verify the transfer of entanglement from a CV to a DV state we illuminate the qubits with the TMS radiation and perform time-resolved qubit measurements. First, we send a square pulse at the pump frequency $\omega_{\rm p}/(2\pi) = \SI{16.805}{\giga\hertz}$ to the JPC, with a normalized amplitude $\epsilon_{\rm p}$ and a duration $t_{\mathrm{pulse}} = \SI{2}{\us} \gg \gamma_{{\rm R},i}^{-1}$ that exceeds the relaxation time of the individual qubits, Fig.~\ref{fig:2}a. Right after the pump pulse, we perform a $\pi/2$ rotation on the qubits to project the measurement along the equator of the Bloch sphere, as shown in Fig.~\ref{fig:2}c. For example, to measure $\langle \sigma_1^{x}\rangle$, we rotate qubit 1  by $\pi/2$ along the $\hat{y}$ axis. For qubit 2, we sweep the measurement axis $\hat{\phi} = \cos\phi\cdot \hat{x} + \sin\phi\cdot \hat{y}$ and extract $\langle \sigma_2^{\phi}\rangle$. Finally, we send a short readout tone of 20 (80) ns to the readout resonator 1 (2), which was optimized to prevent qubit decay during the measurement. 

We repeat the full pulse sequence shown in Fig.~\ref{fig:2}a $10^6$ times and average the results. Here we wait for \SI{5}{\us} in between runs to ensure the qubits have completely decayed to the ground state. Figure~\ref{fig:2}b shows the coherent oscillations expected
for the state $|\Phi\rangle$ in Eq.~\eqref{eq:qubit_state}, for $\phi_{\rm p} = 0$, where the joint expectation value is $\langle \sigma_1^{x}\sigma_2^{\phi}\rangle = \bra{\Phi}\sigma_1^{x}\sigma_2^{\phi}\ket{\Phi} = (N^2 + N)^{\frac{1}{2}}/(N + 1/2) \cos{\phi}$, while $\langle \sigma_1^{x}\rangle = \langle \sigma_2^{\phi}\rangle = 0$ for all angles. This proves phase coherence between the two qubits. We then sweep the phase of the pump tone $\phi_{\rm p}$, which reveals a perfect correlation with the phase of the joint qubit state, as shown in Fig.~\ref{fig:2}c. This shows that the phase of the joint qubit state originates from the phase coherence of the TMS state. 

\begin{figure*}[t]
    \centering
    \includegraphics{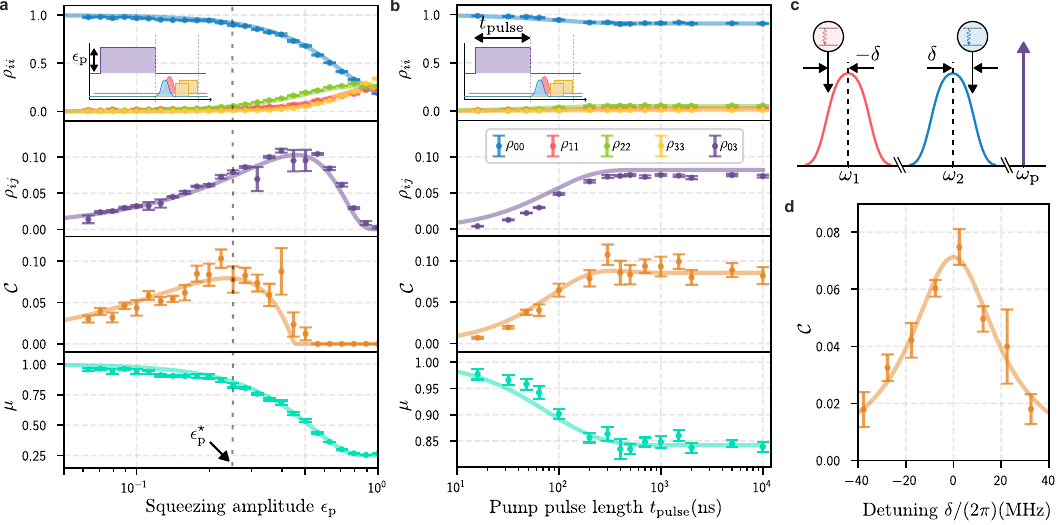}
    \caption{\textbf{Qubit entanglement characterization.} 
    \textbf{a}, Measured real part of the reconstructed density matrix elements $\rho_{ij}$ (see legend in panel b), concurrence $\mathcal{C}$, and purity $\mu$
    for different squeezing amplitudes $\epsilon_{\rm p}$ and \textbf{b}, different duration of the pump pulse $t_{\rm pulse}$. Error bars show the 1$\sigma$ standard error from 7 (panel a) and 10 (panel b) repetitions with $10^7$ averages each. The solid lines correspond to the theory model.
    \textbf{c}, Frequency diagram of the detuning experiment. The two qubits at $\omega_{\mathrm{q}, i}$ are detuned from the center frequencies of the respective TMS modes $\omega_{i}$ by $\delta$, maintaining $\omega_{\rm q,1} + \omega_{\rm q,2} = \omega_{\rm p}$. 
    \textbf{d}, Measured concurrence $\mathcal{C}$ as a function of qubit detuning $\delta$ and a Lorentzian fit with a bandwidth $\delta 
    \omega/2\pi=\SI{44}{\MHz}$ (solid line). Error bars show the 1$\sigma$ standard error from 3 repetitions with $10^7$ averages each.}
    \label{fig:3}
\end{figure*}

\begin{figure*}[t]
    \centering
    \includegraphics{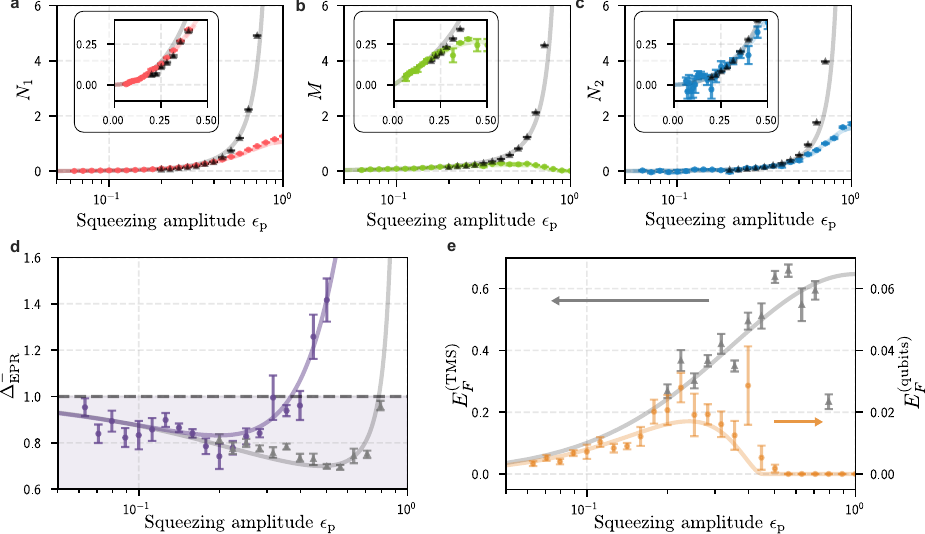}
    \caption{
    \textbf{Photonic characterization and entanglement transfer.} 
    \textbf{a-c}, Average expectation values of the characteristic photon numbers $N_1, M$ and $N_2$ of the TMS state extracted from calibrated linear detection (black triangles) and the measured qubit expectation values (colored circles). Insets show a zoom of the parameter regime where both methods agree. The data is extracted from raw measurements, that is why in the event of noise for qubit 2, $N_2$ can show negative values. 
    \textbf{d}, Measurements of two-mode squeezing where $\Delta_{\mathrm{EPR}}^- <1$ indicates non-separability. Purple data is taken from direct qubit measurements and gray data from calibrated heterodyne detection, as shown in panels (a-c).
  \textbf{e}, Entanglement of formation calculated for the TMS state (grey triangles, left axis) and the joint qubit state (orange circles, right axis). Solid lines are calculated based on the ideal squeezing Hamiltonian model from Eq.~\eqref{eq:HJPC} (grey) and the theory model for the full network, also used in Fig.~\ref{fig:3} (yellow). Error bars show the $1\sigma$ standard error from 7 (5) repetitions with $10^7$ ($2\times10^6$) averages for qubit (linear, 500 kHz) detection.}
    \label{fig:4}
\end{figure*}

Showing correlations in the X and Z basis is only sufficient to certify entanglement if the states are highly correlated ($N \gtrsim 1$) \cite{blinov_observation_2004}, while for weakly correlated states ($N \approx 0.01$), it is necessary to implement full state tomography of the qubit state $\rho$~\cite{steffen_measurement_2006, ryan_tomography_2015}. We extend the measurement sequence to measure in an overcomplete set of 25 different bases (Fig.~\ref{fig:2}d), allowing us to extract the 16 independent expectation values of the two-qubit subspace. For a squeezing amplitude of $\epsilon_{\rm p}^\star = 0.25$, the result is shown in Fig.~\ref{fig:2}e. We then reconstruct $\rho$ from the averaged expectation values $\langle \sigma_1^i \sigma_2^j\rangle$ using a maximum likelihood estimator (MLE) (details in Appendix \ref{app: qubit tomo}), as shown in Fig.~\ref{fig:2}f for the real part of $\rho$. 

In Fig.~\ref{fig:2}e and f, we overlay the measured results (yellow bars) with the theoretical predictions (black frames) 
using the independently extracted parameters from Table~\ref{tab:qubit_params_booktabs} in Appendix~\ref{app: qubit characterization}, which leaves the finite photon transmittances $\eta_i$ between the JPC and the qubits as the only free parameters. We note the excellent agreement between experiment and theory (see Appendix~\ref{app: theory}) for $\eta_i=(0.5, 0.3)$, which is qualitatively consistent with estimates based on the TMS characterization (see Fig.~\ref{fig:4}d and Appendix~\ref{app: TMS characterization}). We also note that the observed coherences $\ket{gg}\bra{ee}$ clearly exceed the unwanted single-qubit diagonal elements $\ket{ge}\bra{ge}$, $\ket{eg}\bra{eg}$, which are due to photon loss $1 - \eta_i$, the bidirectional qubit-waveguide coupling, non-guided qubit decay $\gamma_{\mathrm{ng}, i}$, and qubit decay during the measurement. This is directly observed, i.e., without the need for post-processing of the qubit readout results.

\subsection*{Characterization of qubit entanglement}
In order to study and quantify the entanglement distribution scheme, we vary the JPC pump strength, the pulse lengths, and the qubit frequencies and extract relevant properties of the joint qubit states. In Fig.~\ref{fig:3}a and b, we show (from top to bottom) measurements (symbols) along with theory (lines) of the steady-state qubit populations $\rho_{ii}$, the two-qubit coherence $\rho_{ij}$, the concurrence $\mathcal{C}$, and the purity $\mu=\text{Tr}\{\rho^2\}$ of the two-qubit state, respectively. 

In Fig.~\ref{fig:3}a we sweep the squeezing amplitude $\epsilon_{\rm p}$ and observe a build-up of the two-qubit population, as well as the two-qubit coherence, 
in very good agreement with the theoretical model. 
Similarly, we observe a growing concurrence of up to $\mathcal{C}=0.10 \pm 0.01$ for $\epsilon_{\rm p}\approx \epsilon_{\rm p}^\star$. 
We also calculate the purity $\mu$ of the stabilized two-qubit state, 
which continuously decreases from 1 --- for the vacuum state --- until it saturates at a value of $\mu\approx0.25$. 

For qubits coupled to a unidirectional waveguide~\cite{scigliuzzo_primary_2020, joshi_resonance_2023}, and assuming no other imperfections, their state remains pure, $\mu=1$, for all pump strengths $\epsilon_{\rm p}$ and $\mathcal{C}\rightarrow 1$ when $\epsilon_{\rm p}\rightarrow \infty$. However, for bidirectional waveguides, as used in this work, the emission into uncorrelated left-moving modes prevents the formation of a pure dark state and limits the maximally achievable entanglement to $\mathcal{C}\lesssim 0.26$. This explains the overall reduction of the observed entanglement as well as the appearance of a maximum at finite pumping strength $\epsilon_{\rm p}^\star$. Taking also other sources of imperfections into account, these findings are in excellent agreement with the theoretical expectations, thus demonstrating a complete understanding of the setup and the underlying entanglement formation mechanism.

In Fig.~\ref{fig:3}b, we study the time dependence of this entanglement stabilization mechanism.  
We show the same observables as in panel a, but at a fixed pump strength $\epsilon_{\rm p}^\star$ and as a function of JPC pump pulse duration $t_{\rm pulse}$, which we sweep over three orders of magnitude, from \SI{10}{\nano\second} to \SI{10}{\micro\second}. As a function of $t_{\rm pulse}$, we observe the build up of qubit occupancy, concurrence and the decrease of purity all over a very similar timescale of $t_{\mathrm{stabilize}} \approx \SI{300}{\ns}$, which is approximately the time it takes for the qubits to 
interact with the fields in the waveguide, $\gamma_{\rm R}^{-1}$. 
We show how the state remains stable up to \SI{10}{\us}, which is almost 2 orders of magnitude longer than the stabilization time. We emphasize that fast stabilization is beneficial for further entanglement distillation protocols with high throughput.

Another degree of freedom to quantify this new entanglement distribution scheme is the detuning $\delta$ of the qubits from the respective TMS mode centered at $\omega_i$,  
as sketched in Fig.~\ref{fig:3}c. 
Figure~\ref{fig:3}d shows the measured steady-state concurrence $\mathcal{C}$ of the two-qubit state as a function of $\delta$. Here we see that --- keeping the energy conservation condition $\omega_{\rm q,1} + \omega_{\rm q,2} = \omega_{\rm p}$ fulfilled --- the concurrence decreases with a Lorentzian shape of fitted bandwidth \SI{44}{\MHz}, which is very close to the measured JPC gain bandwidth of $\delta\omega=\SI{46}{\MHz}$ at $\epsilon_{\rm p} = \epsilon_{\rm p}^\star$ (see Appendix~\ref{app: theory}). While this is expected \cite{zippilli_entanglement_2015}, it also highlights an interesting aspect of this work, namely that we can use qubit tomography to directly extract relevant and potentially non-local properties of the itinerant radiation field state. 

\subsection*{Direct verification of Gaussian entanglement}
In the microwave domain, the standard strategy to detect and quantify itinerant fields, such as a TMS state, is to amplify the signals and detect them with a heterodyne measurement setup at room temperature \cite{eichler_characterizing_2012}. We can reconstruct the covariance matrix $V$ of the state at low temperature, i.e. before amplification and losses apply, from the variance of the measured field quadratures $u \in \{I_1, Q_1, I_2, Q_2 \}$. Here, the diagonal elements are given by $V_{ii} = \langle u_i^2 \rangle_{\rm on} - \langle u_i^2 \rangle_{\rm off} + 1/2$, where the subscript indicates that the pump is on or off. When the pump is off, the detected noise $\langle u_i^2 \rangle_{\rm off} = N_{\rm add} + 1/2$ corresponds to the added noise by the amplifiers and the amplified vacuum fluctuations where also the losses between the point of interest and the amplifiers are taken into consideration. The off-diagonal elements come from the covariances of the two modes $V_{ij} = \langle u_i u_j + u_j u_i \rangle_{\rm on}/2$.
To accurately reconstruct the quantum state before it enters the amplification chain (shown by the gray symbols in Fig.~\ref{fig:4}), one must carefully back out the gains $\mathcal{G}_i$ and added noise $N_{\mathrm{add},i}$ of the detection chains, obtained from an independent calibration (see Appendix~\ref{app: TMS characterization}). 
In general, such calibration methods require the use of extra elements inside the fridge, they are frequency dependent, very sensitive to amplifier saturation and cannot capture potential fluctuations of gain and added noise over time.

Here we report an alternative strategy that makes use of qubits as primary detectors \cite{goetz_photon_2017,scigliuzzo_primary_2020}, a method that does not require calibration and can yield higher signal-to-noise ratios for low photon number states, since --- similar to conventional photon detection techniques ---the vacuum noise does not enter the equation (see Appendix~\ref{app: detection}).
To achieve this we use the fact that, in the linear response regime, we can map the photonic operators $a_i$ to the qubit lowering operator $\sigma_{i}^-$, thus obtaining the average photon numbers $N_i$ and the correlations $M$ directly from the steady-state expectation values of the qubits (see Appendix~\ref{app: detection}).

In Fig.~\ref{fig:4}a-c, we compare the estimated TMS photon numbers from qubit measurements with the elements of $V$ reconstructed from independently calibrated heterodyne measurements. We see that both measurements agree at small values of squeezing $\epsilon_{\rm p}$,
where the qubits are in the linear response regime. For photon numbers higher than $\gtrsim 0.2$ the qubit response saturates and the two curves deviate substantially but predictably, following the theory (see Appendix~\ref{app: theory}). We note that the small photon number regime is more easily accessible with the square-law qubit detection compared to the calibrated linear detection due to prohibitive amounts of averaging that would be necessary for the latter. This is particularly apparent for qubit 1 (red curve in panel a) with a quantitative comparison in Appendix~\ref{app: detection}).

With the qubit detection verified, we can now use the Duan-Simon criterion~\cite{duan_inseparability_2000,simon_peres-horodecki_2000},
$\Delta_{\mathrm{EPR}}^- = 1 + N_1 + N_2 - 2M < 1$, to certify entanglement in the TMS state. 
The result is shown in Fig.~\ref{fig:4}d and we find a very good agreement between the values extracted from the qubit detection (purple) and the traditional calibrated linear detection (gray). As expected, the qubit measurement -- due to its saturable nature -- does not capture the full amount of two-mode squeezing obtained at $\epsilon_\textrm{p}\approx 0.55$, as predicted by theory. Importantly, the qubits' measurements provide a lower bound and confirm that the two microwave fields are entangled without the need for noise subtraction or other calibration techniques. 

\subsection*{Quantification of entanglement transfer}
Finally, we also quantify the entanglement transfer~\cite{kraus_discrete_2004} from the original TMS state to the stationary joint qubit state, as shown in Fig.~\ref{fig:4}e. 
To do so, we use the entanglement of formation $E_F$, which is defined both for continuous~\cite{giedke_entanglement_2003, barzanjeh_stationary_2019, flurin_generating_2012} and discrete variable systems~\cite{wootters_entanglement_1998}. 
We observe that $E_F^{(\rm TMS)}$ obtained for the TMS from linear  detection increases as a function of squeezing amplitude, up to $E_F^{(\rm TMS)}\approx0.6$ at $\epsilon_\textrm{p}\approx 0.7$ (gray symbols) before sharply dropping. We attribute this drop to deviations from the ideal non-degenerate amplification Hamiltonian in Eq.~\eqref{eq:HJPC}, which are not included in our model (gray lines).
Similarly, we find that $E_F^{(\rm qubits)}$ obtained from the tomography of the qubit state (yellow symbols) increases with the squeezing amplitude. However, it peaks earlier at $E_F^{(\rm qubits)}\approx0.03$ for $\epsilon_\textrm{p}\approx 0.25$ as expected from the measured squeezing dependence of $\mathcal{C}$ and $\mu$ shown in Fig.~\ref{fig:3}a.

In general, we find that about a tenth of the entanglement of formation present in the TMS state is inherited by the qubits. This is in very good agreement with theory (gray and yellow lines) and can be improved along with maximizing $\mathcal{C}$ as discussed in the Appendix~\ref{app: limitations}. Nevertheless, even in the optimal case, the capacity of a typical CV channel will exceed that of the DV system due to its higher state space. 
When integrating over the bandwidth this becomes even more apparent. In this first experimental realization we distributed a total of up to 36 Mebits/s in the TMS state basis of which up to 12.5 kebits/s is transfered to the joint qubit state, which showcases the advantage of using CV states for the distribution of entanglement to stationary qubits.

\section*{Conclusion}
In summary, we have presented the first experimental realization of the distribution of qubit-qubit entanglement through a correlated photonic reservoir, as originally proposed over two decades ago~\cite{paternostro_complete_2004, kraus_discrete_2004}. This has been possible mainly due to the advances of waveguide QED~\cite{sheremet_waveguide_2023} that allow for very strong and reliable coupling of qubits to propagating waveguide modes. In our proof-of-principle demonstration, the main limitations on the achievable entanglement arose from the bidirectional waveguide couplings, losses in the transmission lines, and unexpectedly large asymmetry in the qubit decay rates. These issues can be readily improved in the next generations of experiments, for example, by terminating the transmission lines, making use of higher-purity~\cite{abdo_teleportation_2025-1}, broadband~\cite{esposito_observation_2022, qiu_broadband_2023}, or multiplexed~\cite{lingua_continuous-variable_2025}, 
squeezing sources and protocols that are currently being developed in the microwave domain. We have also shown --- to the best of our knowledge --- the first calibration-free verification of a two-mode squeezed state in the microwave domain. This has been achieved using the qubits as both photon and correlation detectors. 

In the long run, these new entanglement distribution and measurement schemes open up multiple new avenues for quantum optics experiments and quantum technological applications. Examples include the realization of high-throughput and low control-complexity entanglement distribution that is enabled by the demonstrated fast stabilization time.
The possibility of entangling qubits at very different frequencies in our non-degenerate setting~\cite{sahu_entangling_2023} and the large amount of entanglement contained in broadband CV entangled beams can be directly harnessed for multi-qubit entanglement distribution schemes~\cite{agusti_autonomous_2023} using a single correlated photon source only. The improved detection capabilities will be relevant to unlock a quantum advantage in certain proposed remote sensing scenarios~\cite{barzanjeh_microwave_2020}. Related hybrid CV-DV architectures~\cite{andersen_hybrid_2015}
have already shown promise for quantum sensing~\cite{liu_noise_2022, novikov_hybrid_2025}, measurement-based quantum computation, and the application of non-Gaussian operations and entanglement distillation in the CV domain~\cite{giedke_characterization_2002}. 
\newpage
Note added: While completing this manuscript, we became aware of a recent related cascaded entanglement distribution protocol that does not rely on an entangled reservoir \cite{irfan_autonomous_2025}.

\section*{Acknowledgments}
We thank A. Trioni and C.N. Borja for assistance in device fabrication, C. Siegele for fruitful discussions, IBM for donating the JPC used in this work, and the MIBA machine shop and the ISTA nanofabrication facility for technical support. This work was funded in part by the Austrian Science Fund (FWF) through the excellence cluster quantA 10.55776/COE1 and the SFB BeyondC F7105, as well as the European Union - NextGenerationEU, and ISTA. J.F. and L.K. acknowledge support from the Horizon Europe Program HORIZON-CL4-2022-QUANTUM-01-SGA via Project No.~101113946 OpenSuperQPlus100. J.A. acknowledges support from the QUANTERA project MOLAR with reference PCI2024-153449, funded by MICIU/AEI/10.13039/501100011033 and the European Union. This research is part of the Munich Quantum Valley, which is supported by the Bavarian state government with funds from the Hightech Agenda Bayern Plus.

\appendix
\section{Experimental setup}\label{app: setup}
\subsection*{Sample Fabrication}
We fabricate the qubit sample in a single layer using double-angle shadow evaporation of aluminum. We start with the cleaning of a 10$\times$\SI{10}{\mm^2} high-resistivity silicon chip using an O\textsubscript{2} plasma asher, followed by sonication in hot acetone and isopropanol (IPA) for 10 minutes each at \SI{50}{\celsius}. We dip the chip in buffer hydrofluoric acid for 30 seconds to remove the surface oxide and rinse in water and IPA before spinning the resist. We use a stack of PMMA 950k 4\% on MMA EL 13\% for the undercut of the Dolan bridges. After electron-beam lithography and development, we ion mill the substrate to remove leftovers of the resist and evaporate two layers of 40 and \SI{80}{\nm} of aluminum at a rate of 1 nm/s and do static oxidation at 20 mbar in between to create the junctions. We lift off the unwanted metal in hot dimethyl sulfoxide (DMSO) at \SI{80}{\celsius} and rinse in acetone and IPA for 5 minutes.
\subsection*{Measurement setup}
\begin{figure*}[htbp!]
    \centering
    \includegraphics{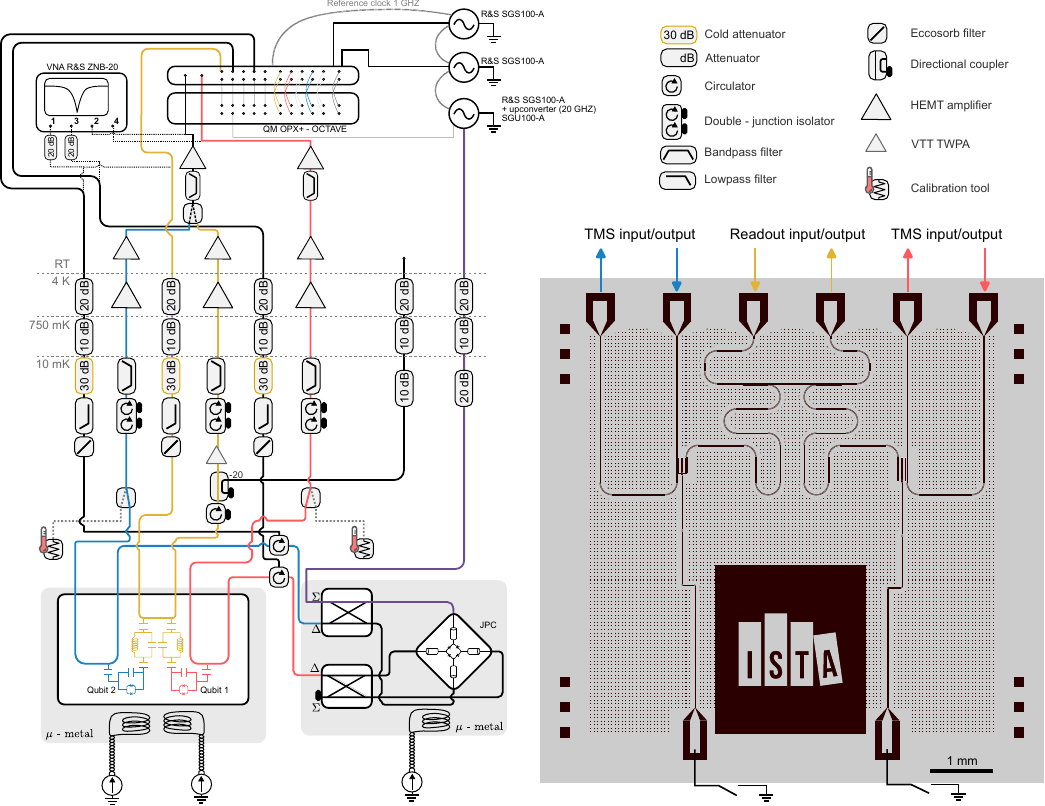}
    \caption{\textbf{Experimental setup.} Cryogenic and room temperature experimental setup together with a CAD image of the qubit chip. A detailed explanation is found in the legend and main text.}
    \label{fig:full_setup}
\end{figure*}

Figure~\ref{fig:full_setup} shows a detailed schematic of the cryogenic and room temperature setup for the experiment. We use a Vector Network Analyzer (VNA) Rohde \& Schwarz ZNB-20 for characterization of the JPC modes, resonators, and qubits through the ports of the waveguides.  
We use superconducting coils attached to the sample boxes to bias the JPC and the qubits, while keeping the on-chip flux lines disconnected. The coils are current-biased using the Delft IVVI rack.
We use \SI{30}{\dB} cryogenic attenuators at the mixing chamber plate to ensure thermalization of the qubit waveguides.
We have connected a Travelling Wave Parametric Amplifier (TWPA), courtesy of VTT, in the resonator readout output line, which is also biased with the IVVI to half-flux to minimize transmission losses, but remained unpumped for the duration of the experiments presented here.

The JPC is connected to two \SI{180}{\degree} commercial hybrids from Krytar.
Both JPC and qubit samples are attached to two separate oxygen-free copper cold fingers and each magnetically shielded by a \textmu-metal shield. We use coaxial copper cables except for the cables connecting the JPC hybrids and the first circulator, which are made of NbTi and help reduce loss between the JPC and the qubits. 

As tools for the calibration of the added noise, we use two cryogenic 50 $\Omega$ loads thermalized to a heater and a RuOx thermometer for precise measurement of the temperature. The load is weakly thermalized to the dilution refrigerator through the two coaxial cables connecting the loads to the switches, resulting in a base temperature of about 200 mK. NbTi coaxial lines connect the output lines from the mixing chamber plate to the three HEMT amplifiers (LNF).

We use RF sources from Rohde \& Schwarz SGS100-A for pumping the JPC and as downconversion LOs for the heterodyne detection. The source that provides the pump tone is combined with an upconverter SGU100-A. The sources are daisy-chained by a \SI{1}{\GHz} reference clock signal.
We use the OPX+ and Octave from Quantum Machines (QM) for generating the pulse sequences. The QM instruments get the clock reference from the R\&S RF sources. The pump tone is triggered by sending a digital pulse from one of the digital ports of the OPX+.

\section{Qubit charaterization}\label{app: qubit characterization}

We bias the flux-tunable transmons $~\SI{300}{\MHz}$ below the flux sweetspot to be resonant with the JPC modes and extract the relevant parameters for the experiment by doing different independent measurements. The parameter values and their respective extraction methods are summarized in Table \ref{tab:qubit_params_booktabs}.

\begin{table*}
\centering
\caption{Qubit chip parameters (in MHz) and photon transmission.}
\label{tab:qubit_params_booktabs}
\begin{tabular}{l c c c c c}
\toprule
\textbf{Parameter} & \textbf{Symbol} & \textbf{Qubit 1} & \textbf{Qubit 2} & \textbf{Extraction method} & \textbf{Fig.} \\
\hline
Qubit transition frequency &$\omega_{\mathrm{q},i}/2\pi$ & 6761 & 10044 & Ramsey sequence & - \\
Qubit anharmonicity &$\alpha_i/2\pi$ & -150 & -180 & Two-tone spectroscopy & - \\
Qubit relaxation rate &$\gamma_{i}/2\pi$ & $1.7 \pm 0.3$ & $1.07 \pm 0.17$ & Pulsed $T_1$ sequence & \ref{fig:qubitstability} \\
Qubit decoherence rate &$\gamma_{\mathrm{d},i}/2\pi$ & $0.89 \pm 0.05$ &$0.57 \pm 0.03$ & Ramsey sequence & \ref{fig:qubitstability} \\
Qubit dephasing rate &$\gamma_{\phi,i}/2\pi$ & $0.04 \pm 0.16$ & $0.03 \pm 0.09$ & $\gamma_{\phi,i} = \gamma_{\mathrm{d},i} - \gamma_{i}/2$ & - \\
Qubit-waveguide coupling &$\gamma_{\mathrm{w},i}/2\pi$ & $1.48$ &$0.7$ & Waveguide spectroscopy & \ref{fig:wQED} \\
Decoherence rate (extracted from waveguide) &$\gamma_{\mathrm{d},i}^{(\mathrm{wQED})}/2\pi$ & $0.83$ & $0.56$ & Waveguide spectroscopy & \ref{fig:wQED} \\
Non-guided relaxation rate &$\gamma_{\mathrm{ng},i}^{(\mathrm{wQED})}/2\pi$ & $0.05$ &$0.35$ & $\gamma_{\mathrm{ng},i} = \gamma_{\mathrm{d},i}^{(\mathrm{wQED})} - \gamma_{\mathrm{w},i}/2 - \gamma_{\phi,i}$ & - \\
Photon transmission & $\eta_i$ & 0.5 & 0.3 & TMS characterization & \ref{fig:TMScharacterization} \\
\hline
\end{tabular}
\end{table*}

\subsection*{Pulsed spectroscopy}
\begin{figure}[t]
    \centering
    \includegraphics{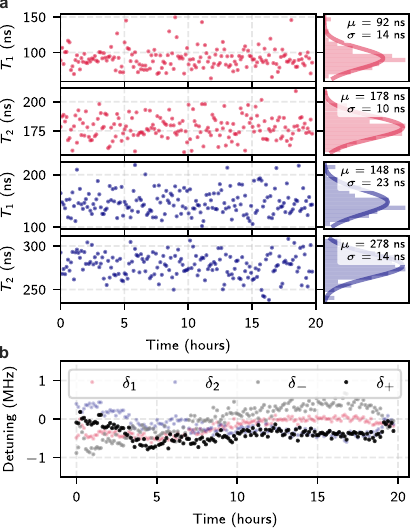}
    \caption{\textbf{Qubit stability.} \textbf{a}, Qubits 1 (red) and 2 (blue) $T_1$ and $T_2$ from energy relaxation and Ramsey measurements during the data acquisition for the squeezing amplitude sweep. \textbf{b}, Qubit detuning during the data acquisition for the squeezing amplitude sweep, where $\delta_i$ is the detuning of qubit $i$ with respect to its corresponding photonic mode and $\delta_\pm = \delta_1 \pm \delta_2$. While $\delta_-$ does not matter for $\delta_- \lesssim \delta\omega \approx 2\pi \cdot \SI{60 }{\MHz}$, a slight drift in $\delta_+ \lesssim \gamma_{\rm w} \approx 2\pi \cdot\SI{1}{MHz}$ reduces the correlations dramatically.}
    \label{fig:qubitstability}
\end{figure}

We interleave $T_1$ and Ramsey measurements in between the measurements of the squeezing amplitude sweep shown in Fig.~\ref{fig:3}. We show the result and compute the mean and the standard deviation in Fig.~\ref{fig:qubitstability}a. From the detuned Ramsey measurement, we track the qubit detuning over time and identify two different contributions to the experiment. It is well known that the correlations of a TMS are maximum when the down-conversion frequencies fulfil the energy conservation relation $\omega_1 + \omega_2 = \omega_{\rm p}$, where $\omega_{\rm p}$ is the frequency of the pump~\cite{zippilli_entanglement_2013}. In our experiment, the down-conversion frequencies are the qubit frequencies. Hence, this condition translated to the qubit detunings reads $\delta_+ = \delta_1 + \delta_2 = 0$ with $\delta_i = \omega_{\mathrm{q},i} - \omega_i$. The range in which $\delta_+$ can vary to still measure correlations is given by the detection bandwidth, in this case $\gamma_{\rm R}$.
On the other hand, the TMS state correlations decrease from the center of the modes due to being filtered by the linewidth of the JPC cavities. This means that if one places the qubits such that $\delta_+ \approx 0$ but detuned from the center frequency of the TMS modes by a detuning $\delta_-/2 = (\delta_1 - \delta_2)/2$, the correlations decrease in a range given by the JPC bandwidth $\delta\omega$, as it is seen in Fig. \ref{fig:3}d.

\subsection*{Waveguide characterization}
\begin{figure}[t]
    \centering
    \includegraphics[width = \columnwidth]{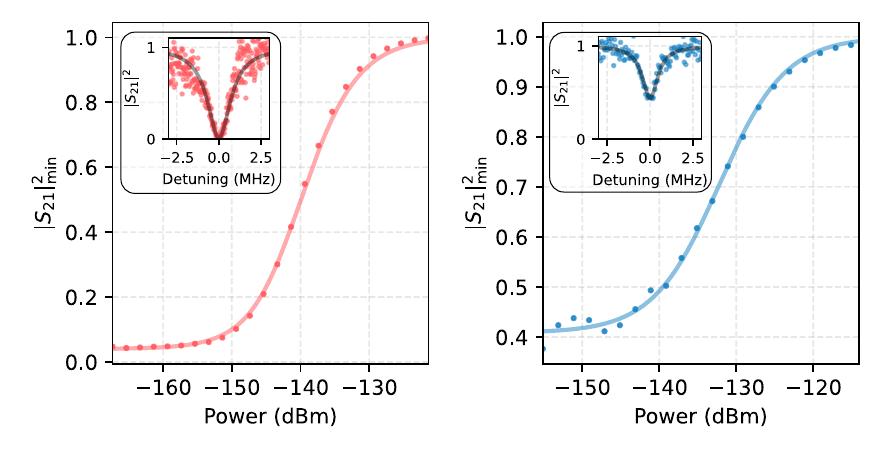}
    \caption{\textbf{Waveguide spectroscopy.} Minimum of transmission as a function of input power for qubit 1 (red) and qubit 2 (blue). Insets: frequency trace at low power and fit to Eq. ~\eqref{eq:S21_wQED}.}
    \label{fig:wQED}
\end{figure}

The transmission through the waveguide depends on the input power and is given by~\cite{sheremet_waveguide_2023}
\begin{equation}{\label{eq:S21_wQED}}
    S_{21}(\Delta) = 1 - \frac{\gamma_{\mathrm{w},i}}{2\gamma_{\mathrm{d},i}}\frac{1 - i\Delta/\gamma_{\mathrm{d},i}}{1 + (\frac{\Delta}{\gamma_{2,i}})^2 + \frac{\Omega^2}{\gamma_{\mathrm{w},i}\gamma_{\mathrm{d},i}}},
\end{equation}
where $\gamma_{\mathrm{w},i}$ is the qubit-waveguide coupling, $\gamma_{\mathrm{d},i} = \gamma_{\mathrm{w},i}/2 + \gamma_{\phi,i} + \gamma_{\mathrm{ng},i}$ the total decoherence rate of qubit $i$, $\Delta$  the drive detuning and $\Omega$ the drive amplitude. At low drive amplitudes $\Omega \ll \gamma_{\mathrm{d},i}$, we fit the response to a Lorentzian lineshape from where we extract both $\gamma_{\mathrm{d},i}$ and the ratio $\gamma_{\mathrm{w},i}/(2\gamma_{\mathrm{d},i})$.
The result of these fits is shown in the insets of Fig.~\ref{fig:wQED}. 

We then sweep the drive strength and fit the complex response to Eq.~\eqref{eq:S21_wQED} to extract the minimum of the transmission for the different powers. In Fig.~\ref{fig:wQED} we see good agreement between the measured values and the theoretical fit, from which we extract the attenuation of $A = [ 122.2 \pm 0.1, 131.4 \pm 0.1 ]$ dB. 
Note that while the transmission for qubit 1 is almost completely suppressed on resonance with a residual transmission that can be and be attributed to qubit dephasing, for qubit 2, the shallow dip indicates significant losses into other non-guided modes, i.e., channels different than the waveguide. This could be caused by a neighboring two-level system (TLS) or resonances of the resonator transition inside the transmon well with higher-order qubit transitions.

\section{Two-qubit tomography}\label{app: qubit tomo}
We perform tomography of the qubit state by measuring the qubits in different basis combinations via the dispersively coupled resonators. Given the short qubit relaxation time $T_1 \sim \SI{100}{ns}$, we optimize for a short tomography sequence (shown in Fig.~\ref{fig:2}d) that minimizes state preparation and measurement errors. We prepare short \SI{16}{\ns} Gaussian-shaped pulses optimized using DRAG~\cite{chen_measuring_2016} to not leak out of the qubit subspace. Then, we send a readout tone of 20 (80) ns to the readout resonator 1 (2), compromising the readout fidelity to avoid qubit decay. The response of the resonator is projected along the axis that separates the two resonator responses and integrated using a matched filter that maximizes the separation between states $\ket{g}$ and $\ket{e}$. Although the single-shot readout fidelity is 60 (57)\%, the rapid decay of the qubits allows for a fast repetition rate close to \SI{200}{\kHz}. 

\begin{figure*}[t]
    \centering
    \includegraphics{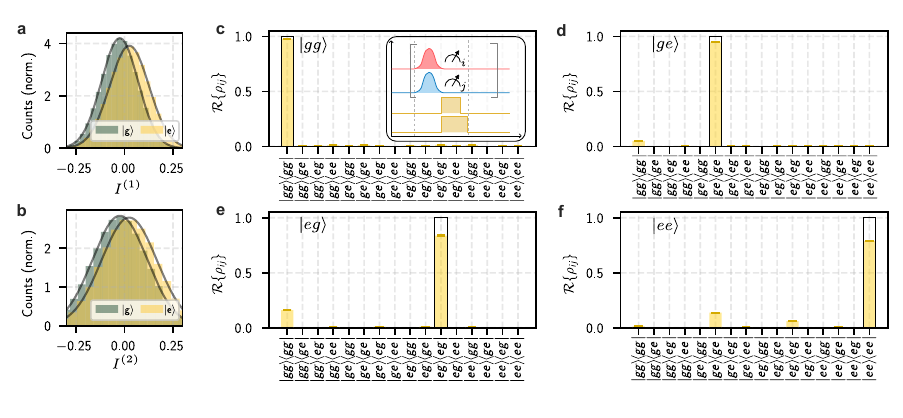}
    \caption{\textbf{Two-qubit tomography.} \textbf{a}, IQ blobs from dispersive readout of qubit 1 and \textbf{b}, qubit 2. \textbf{c-f}, Calibration of the tomography with the basis states $\ket{gg}$, $\ket{ge}$, $\ket{eg}$, and $\ket{ee}$. Black lines indicate the target state.}
    \label{fig:tomocharacterization}
\end{figure*}

Before sending the TMS signal, we measure the averaged readout for having prepared the qubits in the ground state and along the equator of the Bloch sphere, and scale the readout response as 
\begin{equation}
    I^{(i)} = \frac{\tilde{I}^{(i)} - I_{x}^{(i)}}{I_g^{(i)} - I_x^{(i)}},
\end{equation}
where $I_g^{(i)}, I_x^{(i)}$ correspond to the averaged value of $I^{(i)}$ when preparing the states $\ket{g}, (\ket{g} + \ket{e})/\sqrt{2}$ in qubit $i$. 
Then, we measure in any combination of the basis ${Z, +X, -X, +Y, -Y}$ in both qubits, which gives a total of 25 different measurement bases.
We average over the measurement axis that are equivalent, such as $IX,-IX$, to eliminate offsets coming from imperfect rotations and show only the averaged value of the 16 independent expectation values.
We apply a rotation transformation to the expectation values in post-processing to rotate the reference frame of qubit 2 such that it is aligned with that of qubit 1. The rotation angle $\phi$ can be extracted  from 
\begin{equation}
    \tan \phi =\frac{1}{2} \bigg( \frac{\langle XY \rangle}{\langle XX \rangle} - \frac{\langle YX \rangle}{\langle YY \rangle} \bigg).
\end{equation}
We repeat the full tomographic sequence a total of $10^7$ times and estimate the density matrix $\rho$ that most likely gives this set of averaged expectation values using the convex optimization Python library \textit{cvxpy}. It finds $\rho$ such that it minimizes the cost function
\begin{equation}
    \sum_{k = 1}^{16} |\mathrm{Tr}\{\rho O_k \} - \langle O_k \rangle|^2,
\end{equation}
while imposing the constraints that: $\rho$ is positive semidefinite, Hermitian and $\mathrm{Tr}\{\rho\} = 1$. 
With this procedure and at the beginning of each of the experiments we calibrate the readout by preparing the four basis states $\{\ket{gg}, \ket{ge}, \ket{eg}, \ket{ee} \}$ and obtain an average fidelity of 97.4, 94.0, 84.0, and 78.7\% respectively, which agree with the expected fidelity due state preparation and measurement errors due to the short qubit lifetimes.
Error bars throughout the article show the standard error from \numrange{7}{10} independent measurements. 

\section{Two-mode squeezed state characterization}\label{app: TMS characterization}
\begin{figure*}[t]
    \centering
    \includegraphics{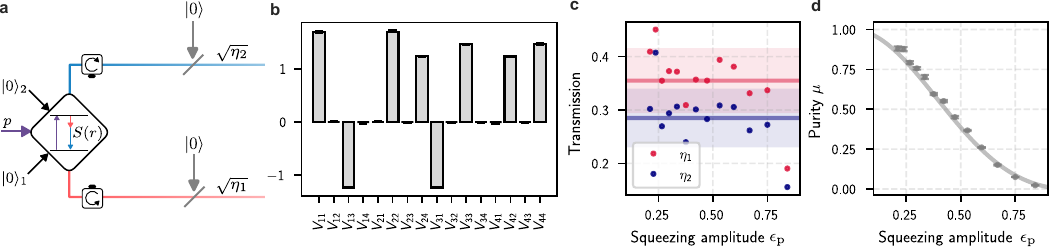}
    \caption{\textbf{TMS characterization.} \textbf{a}, Sketch of the TMSV model where losses are modeled with a partially transmitting beamsplitter that mixes the TMS state with the vacuum state. \textbf{b}, Experimental covariance matrix (bars) corresponding to a squeezing amplitude of $\epsilon_{\rm p} = 0.6$ and fit to the TMSV model from Eq.~\eqref{eq:AppCovarianceMatrix} (black boxes). \textbf{c}, Fitted transmission $\eta_1$, $\eta_2$ for the different squeezing amplitudes. Lines indicate the mean, and shaded areas are one standard deviation. \textbf{d}, Calculated purity $\mu$ of the TMS state.} 
    \label{fig:TMScharacterization}
\end{figure*}

The TMS state is a Gaussian state; therefore, it can be described, without loss of generality, by the covariance matrix of the field quadratures. Assuming a model for a two-mode squeezed vacuum (TMSV) state with an ideal squeezing Hamiltonian as in Eq.~\eqref{eq:HJPC} and a setup as shown in Fig.~\ref{fig:TMScharacterization}a, the covariance matrix reads~\cite{abdo_teleportation_2025-1} 
\begin{equation}\label{eq:AppCovarianceMatrix}
V = 
\begin{pmatrix}
V_{11} & 0 & V_{13} & 0 \\
0 & V_{11} & 0 & -V_{13} \\
V_{13} & 0 & V_{33} & 0 \\
0 & -V_{13} & 0 & V_{33}
\end{pmatrix},
\end{equation}
with
\begin{subequations}
\begin{align}
    &V_{11} = \frac{\eta_1 \cosh (2r) + (1 - \eta_1)}{2},\\
    &V_{33} = \frac{\eta_2 \cosh (2r) + (1 - \eta_2)}{2},\\
    &V_{13} = \frac{\sqrt{\eta_1 \eta_2} \sinh (2r)}{2} .
\end{align}
\end{subequations}
In our case, the covariance matrix is modeled in terms of the transmission in the channels $\eta_1, \eta_2$ (see Fig.~\ref{fig:TMScharacterization}a) and the squeezing parameter $r$.
By fitting the covariance matrix after having subtracted the amplifier noise, we can get an estimate of the photon loss $1 - \eta_i$, which takes into account both attenuation in the cables, connectors, and different components as well as internal losses in the JPC and imperfect two-mode squeezing generation. We find that it is independent of the squeezing amplitude up to the last value, where the TMSV model starts to fail. We can take these values for $\eta_i$ as a lower bound for the actual transmission, while we find that the values $\eta_i = (0.5, 0.3)$ reproduce better the experimental data. We also use these measurements to match the applied pump power $P_{\rm p}$ to the dimensionless parameter $\epsilon_{\rm p}$ related by 
\begin{equation}
    \epsilon_{\rm p} = 10^{(P_{\rm p} - \alpha)/20},
\end{equation}
where $P_{\rm p}$ is in units of dBm, and we find a value of $\alpha = -46$ dBm. 

\begin{figure}[t]
    \centering
    \includegraphics[width=\columnwidth]{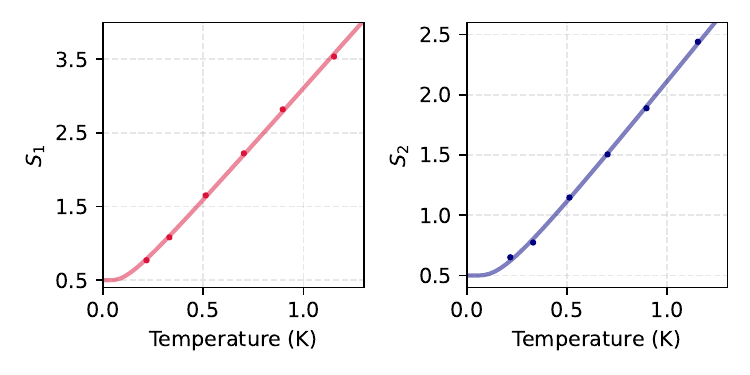}
    \caption{\textbf{Added noise calibration.} Calibration of $N_{\rm add}$ from fitting the normalized power spectral density $S_i$ to Eq.~\eqref{eq: nadd}. The thermal noise is generated using a heated \SI{50}{\Omega} load (shown in Fig.~\ref{fig:full_setup}).}
    \label{fig:temp_calibration}
\end{figure}

We measure the two output signals of the JPC using the down-conversion boards inside the OPX+ and Octave. We digitize the in-phase quadrature $ I = (a_{\rm out} + a_{\rm out}^\dagger)/\sqrt{2}$, with $a_{\rm out}$ the annihilation operator of the output field,  and digitally rotate the phase to obtain $ Q = -i(a_{\rm out} - a_{\rm out}^\dagger)/\sqrt{2}$ of each mode in intervals of \SI{2}{\us} at a sampling rate of \SI{250}{\MHz} and integrating the signal in time with a Chebyshev finite input response (FIR) filter at cutoff frequency $\omega_c/(2\pi) = \SI{500}{\kHz}$. We acquire a total of $10^7$ measurement records for each value of $\epsilon_{\rm p}$. The local oscillator (LO) frequencies used for the down-conversion are referenced using a \SI{1}{\GHz} clock signal and fulfill the condition $ \omega_{\mathrm{LO},1}+\omega_{\mathrm{LO},2}=\omega_{\rm p}$.
Hence, for one channel, we down-convert in the positive frequency spectrum, whereas for the other, we use the negative frequency spectrum. We correct for this by transforming the quadratures of one of the channels, like
$
    (I_2,Q_2) \longrightarrow (Q_2,I_2)
$.
To obtain the correct output fields, we must normalize the detected signal with respect to the amplitude of the vacuum fluctuations, as the amplification chain adds noise. 
We assume that when the pump is off, we are measuring the vacuum state; therefore, at room temperature, we are measuring a total of $N_{\mathrm{add},i}+1/2$ photons in each of the modes $i=1,2$. The mean value of the quadratures $\Bar{u}\in \{I_1,Q_1,I_2,Q_2 \}$ is null, and the variance corresponds to $\langle \bar{u}^2 \rangle$. We calculate the scale factor
\begin{equation}
    \zeta_i = \sqrt{\frac{2(N_{\mathrm{add},i}+1/2)}{\langle I_{\mathrm{add},i}^2\rangle + \langle Q_{\mathrm{add},i}^2 \rangle}},
\end{equation}
using the variances of the vacuum state (pump off), and obtain the normalized variances $ u = \Bar{u}\cdot\zeta_i$ for the signal (pump on).
We estimate the number of added photons by using a calibration tool comprised of a $\SI{50}{\Omega}$ load that we heat up together with a thermometer that measures the local temperature. The detected noise at frequency $\omega_i$ as a function of temperature is 
\begin{equation}\label{eq: nadd}
    N_i = \hbar \omega_i \cdot \mathrm{RBW} \cdot \mathcal{G}_i \cdot \bigg[\frac{1}{2} \mathrm{coth}\bigg(\frac{\hbar\omega_i}{2k_BT}\bigg) + N_{\mathrm{add},i}\bigg].
\end{equation}
In Fig.~\ref{fig:temp_calibration}, we show the measured noise density in units of number of photons $S_i = N_i/(\hbar\omega_i \cdot \mathrm{RBW} \cdot \mathcal{G}_i) - N_{\mathrm{add},i}$ where have have extracted values for the gain in the two modes $(\mathcal{G}_1, \mathcal{G}_2) = (108.67 \pm 0.06, 102.05 \pm 0.07)$ dB and added noise of $(N_{\mathrm{add},1}, N_{\mathrm{add},2}) = (26.8 \pm 0.4, 13.40 \pm 0.2)$. 

\section{Theoretical model}\label{app: theory}
We consider the setup shown in Fig.~\ref{fig:1} of the main text and assume that the transmission lines connecting the outputs of the JPC with the qubits have a linear dispersion relation and that propagation delays can be neglected. Under these conditions, we can adiabatically eliminate the modes of the transmission lines and model the system in terms of a master equation for the reduced state of the JPC and the qubits,
\begin{equation}\label{eq:AppFullModel}
	\dot{\rho}=(\mathcal{L}_{\rm JPC}+\mathcal{L}_{\rm q}+\mathcal{L}_{\rm int})\rho.
\end{equation}
The first term describes the JPC, which we model in terms of two cavity modes with frequencies $\omega_1$ and $\omega_2$, which are coupled through an externally driven down-conversion process. By moving into an interaction picture with respect to $H_0=\sum_{i=1,2}\omega_{i}a_i^\dagger a_i$ and assuming $\omega_1+\omega_2=\omega_{\rm p}$, where $\omega_{\rm p}$ is the external driving frequency, this process is described by the two-mode squeezing Hamiltonian
\begin{equation}
	H_{\rm JPC}=i\frac{\sqrt{\kappa_1\kappa_2}}{2}\epsilon_{\rm p} (e^{i \phi_{\rm p}}a_1^\dagger a_2^\dagger - e^{-i \phi_{\rm p}}a_1 a_2 ). 
\end{equation}
Here, $\kappa_i$ denotes the decay rates of the cavity modes and $\epsilon_{\rm p}\in[0,1)$ is the dimensionless driving strength. By including the decay of the modes into the transmission lines, we obtain 
\begin{equation}
	\mathcal{L}_{\rm JPC}\rho = -i[H_{\rm JPC},\rho] + \sum_{i=1,2}\kappa_i \mathcal{D}[a_i]\rho,
\end{equation}
where $\mathcal{D}[c]\rho=c\rho c^\dag - (c^\dag c\rho+\rho c^\dag c)/2$. 
The qubits, described by the second line in Eq.~\eqref{eq:AppFullModel} are detuned from the respective cavities by $\delta_i=\omega_{{\rm q},i}-\omega_i$ and decay into the right- and left-propagating modes of the transmission lines with rates $\gamma_{R,i}$ and $\gamma_{L,i}$, respectively. Therefore, we model the dynamics of the undriven qubits by 
\begin{equation}
    \begin{split}
	\mathcal{L}_{\rm q}\rho =& -i[H_{\rm q},\rho] + \underbrace{\gamma_{R,1} \mathcal{D}[\sigma^-_1]\rho + \gamma_{R,2} \mathcal{D}[\sigma^-_2]\rho}_{\text{decay to the right}} \\
    &+ \underbrace{\gamma_{L,1} \mathcal{D}[\sigma^-_1]\rho + \gamma_{L,2} \mathcal{D}[\sigma^-_2]\rho}_{\text{decay to the left}} \\
    &+ \underbrace{\frac{\gamma_{\phi,1}}{2} \mathcal{D}[\sigma^z_1]\rho + \frac{\gamma_{\phi,2}}{2} \mathcal{D}[\sigma^z_2]\rho}_{\text{dephasing}} \\
    &+\underbrace{\gamma_{\rm ng,1} \mathcal{D}[\sigma^-_1]\rho + \gamma_{\rm ng,2} \mathcal{D}[\sigma^-_2]\rho}_{\text{decay into non-guided modes}}.
    \end{split}
\end{equation}
Here, $H_{\rm q}=\sum_i \delta_i \sigma^z_i/2$ and for each qubit we have included dephasing with rate $\gamma_{\phi,i}$ and a decay with rate $\gamma_{\rm{ng},i}$ into other non-guided modes. Note that in the current experiment setup the decay of the qubits into the transmission line is fully symmetric, i.e., $\gamma_{L,i}=\gamma_{R,i}=\gamma_{\mathrm{w}, i}/2$, but by setting $\gamma_{L,i}\ll \gamma_{R,i}$, the same theoretical framework can be used to model directional qubit-waveguide interactions as well. 

The JPC and the qubits are separated by a circulator, such that the driving of the qubits by the JPC can be described by a cascaded interaction of the form 
\begin{equation}
\begin{split}
	\mathcal{L}_{\rm int}\rho=&\sqrt{\kappa_1\gamma_{R,1} \eta_1} \left([a_1\rho,\sigma^+_1]+[\sigma^-_1,\rho a^\dagger_1]\right)\\
    &+\sqrt{\kappa_2\gamma_{R,2} \eta_2} \left([a_2\rho,\sigma^+_2]+[\sigma^-_2,\rho a^\dagger_2]\right).
\end{split}
\end{equation}
Here, we have introduced the parameters $\eta_i\le 1$ to account for all photon losses that occur between the JPC and the qubits. 
\subsubsection*{Effective Qubit Master Equation}
In the limit $\kappa_{1},\kappa_2 \gg \gamma_{{\rm w}, 1},\gamma_{{\rm w}, 2}$, also the dynamics of the JPC can be eliminated using a Markov approximation and we obtain an effective master equation for the reduced stated of the qubits, $\rho_{\rm q}={\rm Tr}_{\rm JPC}\{\rho\}$. This master equation reads
\begin{equation}\label{eq:App_Qubit_ME}
\begin{split}
        \dot{\rho}_{\rm q}=&-i[H_{\rm q},\rho_{\rm q}]\\
        &+(N_{\rm ph,1}+1)\gamma_{R,1}\mathcal{D}[\sigma^-_1]\rho_{\rm q}+(N_{\rm ph,2}+1)\gamma_{R,2}\mathcal{D}[\sigma^-_2]\rho_{\rm q}\\
        &+N_{\rm ph,1}\gamma_{R,1}\mathcal{D}[\sigma^+_1]\rho_{\rm q}+N_{\rm ph,2}\gamma_{R,2}\mathcal{D}[\sigma^+_2]\rho_{\rm q}\\
        &+(\gamma_{\rm ng,1}+\gamma_{L,1})\mathcal{D}[\sigma_1^-]\rho_{\rm q}+(\gamma_{\rm ng,2}+\gamma_{L,2})\mathcal{D}[\sigma_2^-]\rho_{\rm q}\\
        &+\frac{\gamma_{\phi,1}}{2}\mathcal{D}[\sigma_1^z]\rho_{\rm q}+\frac{\gamma_{\phi,2}}{2}\mathcal{D}[\sigma_2^z]\rho_{\rm q}\\
        &+\sqrt{\gamma_{R,1}\gamma_{R,2}}M_{12}[\sigma^+_1,[\sigma^+_2,\rho_{\rm q}]]\\
        &+\sqrt{\gamma_{R,1}\gamma_{R,2}}M^*_{12}[\sigma^-_1,[\sigma^-_2,\rho_{\rm q}]],
\end{split}
\end{equation}  
where we have defined $N_{\rm ph,i}=\eta_i\langle a_i^\dag a_i\rangle$ and $M_{12}= \sqrt{\eta_1\eta_2}\langle a_1 a_2\rangle$. For the current experimental setup, the condition for eliminating the JPC modes is well justified, and all numerical results shown in the main text are derived from the steady state of Eq.~\eqref{eq:App_Qubit_ME} for $\gamma_{R,i}=\gamma_{L,i}=\gamma_{{\rm w},i}/2$. All the other parameters are extracted from independent measurements of the qubits and the JPC and are summarized in Table~\ref{tab:qubit_params_booktabs}. 

\subsubsection*{Gain-bandwidth relation}
While the output of the JPC drives the qubits, the qubits see a number of correlated photons within a certain bandwidth $\delta\omega$, which depends on the gain of the parametric amplifier. For a parametric amplifier, the gain is given by~\cite{flurin_generating_2012}
\begin{equation}       
    G_0=\cosh(r)^2=\left(\frac{\epsilon_{\rm p}^2+1}{\epsilon_{\rm p}^2-1}\right)^2 = \left(\frac{P+1}{P-1}\right)^2,
\end{equation}
\noindent where we have defined pump power $P=\epsilon_{\rm p}^2$. The amplification bandwidth is then defined as the bandwidth where the gain reduces by \SI{3}{\dB},
\begin{equation}
    G(\delta \omega)=\frac{1}{2}G(0).
\end{equation}
We define the full-width at half maximum (FWHM) $\delta \omega$ as~\cite{metelmann_quantum-limited_2025}
\begin{equation}
    \delta \omega = \frac{\kappa}{\sqrt{G_0}+1} \approx  \frac{\kappa}{\sqrt{G_0}} = \kappa \left(\frac{P-1}{P+1}\right)^2.
\end{equation}
We can then use $\delta \omega$, which is experimentally accessible, to obtain $\kappa$, which is necessary for the theoretical model. Taking the experimental values $\kappa/(2\pi)=\SI{60}{\mega\hertz}$ and $P=\epsilon_{\rm p}^2=0.0625$, we obtain $\delta \omega/(2\pi) \approx \SI{46}{\mega\hertz}$, in agreement with the experimentally observed amplification bandwidth in Fig.~\ref{fig:3}d.

\section{Entanglement measures}\label{app: entanglement}
Here, we present the entanglement measures we report in the main text.
\subsection*{Continuous-variable entanglement}
The output of the JPC is fully characterized by the covariance matrix $V$ in Eq.~\eqref{eq:AppCovarianceMatrix}.
Its main properties can be reduced to its symplectic eigenvalues $\nu_{\pm}$~\cite{adesso_gaussian_2005}, given by 
\begin{equation}
    \nu_\pm(r)=\sqrt{\frac{\Delta\pm\sqrt{\Delta^2-4 \text{det}(V)}}{2}}.
\end{equation}
\noindent with $\Delta=V_{11}^2+V_{33}^2-2 V_{13}^2$. The smallest symplectic eigenvalue $\nu_-$ can then be used to calculate the entanglement of formation (EOF), given by
\begin{equation}
    E_F=\text{max}\{0,h(2\nu_-)\},
\end{equation}
\noindent with $h(x)=\frac{(1+x)^2}{4x}\log_2{\Big[\frac{(1+x)^2}{4x}\Big]}-\frac{(1-x)^2}{4x}\log_2{\Big[\frac{(1-x)^2}{4x}\Big]}$.
The EOF quantifies the number of pure single states, i.e., EPR pairs, that are needed on average to generate the state through local operations and classical communication (LOCC) only.

\subsubsection*{Duan-Simon criterion}
Alternatively, to demonstrate entanglement between the two modes, we can apply the Duan-Simon criterion~\cite{duan_inseparability_2000, simon_peres-horodecki_2000}. We define the non-local observables 
\begin{equation}
    X_{\pm}(\varphi) = (I_1(\varphi)\pm I_2)/\sqrt{2},
\end{equation}
\begin{equation}
    P_{\pm}(\varphi) = (Q_1(\varphi)\pm Q_2)/\sqrt{2},
\end{equation}
where $\varphi$ rotates the quadratures of one of the modes.
Then, Duan-Simon criterion states that if the inequality
\begin{equation}
    \Delta^{-}_{\rm EPR}= (\Delta X_-)^2 + (\Delta P_+)^2 \ge 1,
\end{equation}
is violated, then the two modes are entangled. Thus, measuring $\Delta_{\rm EPR}<1$ would certify the existence of quantum correlations between the two modes. We can calculate the variance of the non-local observables by 
\begin{equation}
    \Delta(O_{\pm}) = \frac{1}{2}(\langle O_{\pm}^2\rangle - N_{\mathrm{add},1} - N_{\mathrm{add},2}),
\end{equation}
where $O=X,P$. With the joint quadratures, we calculate the variances by subtracting the added number of photons.

\subsubsection*{Purity}
For a complete description of the TMS, its purity $0 \le \mu \le 1$ can be estimated from the covariance matrix $V$ as 
\begin{equation}
    \mu=\frac{1}{4 \sqrt{\text{det}(V)}}.
\end{equation}
The lower bound $\mu=0$ is achieved for a completely mixed two-mode state, while $\mu=1$ would certify that the state is in a pure state $|\Psi_{\rm TMS}\rangle$. This is shown in Fig.~\ref{fig:TMScharacterization}d. A pump strength around $\epsilon_{\rm p}\approx 0.5$ reduces the purity to $\mu\approx 1/3$, which is the minimal purity to generate qubit-qubit entanglement~\cite{agusti_long-distance_2022}. This value agrees with the experimental data from Fig.~\ref{fig:3}.

\subsection*{Discrete-variable entanglement}

Quantifying entanglement for a discrete-variable system differs substantially from a continuous-variable system. Here, assuming a bipartite state characterized by a two-qubit density matrix $\rho_{\rm q}$, we can use the concurrence $\mathcal{C}$ as an entanglement monotone~\cite{hill_entanglement_1997, wootters_entanglement_1998, horodecki_quantum_2009}. It is defined as
\begin{equation}
    \mathcal{C}(\rho_{\rm q})=\text{max}(0,\lambda_1-\lambda_2-\lambda_3-\lambda_4),
\end{equation}
where $\lambda_i \ge 0$ are the eigenvalues in descending order of $R=\sqrt{\sqrt{\rho}\tilde{\rho}\sqrt{\rho}}$ and $\tilde{\rho}=(\sigma_y\otimes \sigma_y)\rho^*(\sigma_y\otimes \sigma_y)$ is the spin-flipped state with $*$ indicating the conjugate. The concurrence is zero for any separable state, $\mathcal{C}(\rho_{\rm sep})=0$, and is maximal for any of the Bell states, i.e., 
$\mathcal{C}(\rho_{\rm Bell})=1$. Alternatively, we can also define the EOF, similar to the continuous-variable case. For a two-qubit system, the EOF is closely related to the concurrence~\cite{hill_entanglement_1997},
\begin{equation}
    E_F=h\left(\frac{1+\sqrt{1-\mathcal{C}^2}}{2}\right),
\end{equation}
where $h(x)=-x \log_2{(x)}-(1-x) \log_2{(1-x)}$ is the Shannon entropy function. Similar to the continuous-variable case, it quantifies how much entanglement (in  ebits) is necessary, on average, to prepare such a state.

\subsubsection*{Bidirectional vs. unidirectional channels}
In a scenario where the qubits are coupled to a single TMS reservoir and assuming also otherwise ideal conditions with $N_{\rm ph,i}=\sinh^2(r)$ and $M_{12}=\cosh(r)\sinh(r)e^{i\phi_{\rm p}}$, it can be shown that the concurrence of the steady state is given by 
\begin{equation}
\mathcal{C}(r)= \tanh{(2r)},
\end{equation} 
and can reach a value of $\mathcal{C}\approx 1$ for sufficiently strong squeezing. In the current experimental setup, this is not the case since the coupling to the left-propagating modes in the transmission line introduces an additional decay channel for each qubits into an uncorrelated reservoir. By assuming $\gamma_{L,i}=\gamma_{R,i}=\gamma_{\rm w}/2$, and otherwise ideal conditions as before, we can evaluate the maximally achievable concurrence $\mathcal{C}^\star$ for the bidirectional scenario. This upper bound can be obtained analytically,
\begin{equation}
    \mathcal{C}^\star=\frac{13\sqrt{13}-19}{108}\approx 0.258,
\end{equation}
and is reached at an optimal squeezing strength of 
\begin{equation}
    r^\star=\frac{1}{2}\log{\left(\frac{4+\sqrt{13}}{3}\right)}=\tanh^{-1}{\left(\frac{\sqrt{13}-1}{6}\right)}\approx 0.465.
\end{equation}
This value corresponds to an optimal pump strength $\epsilon_{\rm p}^\star=\tanh{(r^\star/2)}\approx 0.22$ for our TMS Hamiltonian. 
Any asymmetries, e.g., $\gamma_{\rm w, 1}\neq \gamma_{\rm w, 2}$, will further reduce this upper bound, which explains the comparatively low observed values of $\mathcal{C}$ when compared to the ideal, unidirectional scenario. The upper bound obtained for the concurrence also limits the amount of achievable entanglement of formation, which is given by $E^{\star}_{F}\approx 0.12 $. 

\section{Comparison between photonic detection schemes}\label{app: detection}

Here, we offer a detailed analysis of how one can use the qubit as a probe of TMS states. To do so, we evaluate the signal-to-noise ratio (SNR) for the standard amplified heterodyne detection and compare it with using qubits as detectors. We calculate the expected single-shot variance for the two measurements and compare them.

\subsection*{Room-temperature heterodyne detection}
Consider a single input mode $a_{\rm in}$ that we want to amplify. An amplification chain of total gain $\mathcal{G}$ produces as an output field mode 
\begin{equation}
    a_{\rm out} = \sqrt{\mathcal{G}} a_{\rm in} + \sqrt{\mathcal{G} - 1}h_{\rm in}^\dagger,
\end{equation}
where $h_{\rm in}$ represents a thermal noise field that is added during amplification, with $N_{\rm add} = \langle h_{\rm in}^\dagger h_{\rm in} \rangle $. The output field is then mixed with a local oscillator of frequency $\omega_{\rm LO}$ and down-converted to the two quadratures of the incoming field 
\begin{equation}
        I = (a_{\rm out} + a_{\rm out}^\dagger)/\sqrt{2},
\end{equation}
\begin{equation}
        Q = -i(a_{\rm out} - a_{\rm out}^\dagger)/\sqrt{2}.
\end{equation}
We want to estimate the covariance matrix, or equivalently, the number of photons in one of the thermal fields. As an example for mode 1
\begin{equation}
    N_1 = V_{11} - 1/2 - N_{\rm add}.
\end{equation}
If $V_{11} = \sigma_{11}^2$, the variance is~\cite{sahu_entangling_2023}
\begin{equation}
    \mathrm{Var}\{ V_{11} \} = \frac{\sigma_{11}^4}{N - 1},
\end{equation}
with $N$ the number of samples. It then simplifies to
\begin{equation}
    \mathrm{Var}\{ N_{1} \} = \frac{(1/2 + N_{\rm add} + N_1)^2}{N - 1}.
\end{equation}
If we define the signal-to-noise ratio as
\begin{equation}
    \mathrm{SNR} = \frac{N_1}{1/2 + N_{\rm add}}.
\end{equation}
The single-shot variance can be rewritten as a function of $N_1$ and the SNR as 
\begin{equation}
    \mathrm{Var}\{ N_{1} \} = N_1^2(1 + \mathrm{SNR}^{-1})^2.
\end{equation}
We find that the variance depends on $N_1$, but for small values of $N_1$ it saturates to a value determined by the added photon number $N_{\rm add}$ and it is fundamentally bounded by the vacuum noise.

\subsection*{Qubit detection}
To estimate the average photon number of the thermal distribution from the steady-state qubit expectation values, we assume that the TMS driving the qubits is weakly populated, $N_i<1$. This allows us to linearize the qubit $\sigma_i^- \rightarrow b_i$ with $b_i$ being a bosonic operator. The qubit excitation probability in the steady state can be calculated from~\cite{agusti_long-distance_2022}
\begin{equation}\label{eq: integral P_e}
    P_{e,i} = \langle b_i^\dagger b_i \rangle = \int_{-\infty}^{+\infty} \frac{d\omega}{2\pi} \int_{-\infty}^{+\infty} \frac{d\omega^\prime}{2\pi} e^{i (\omega - \omega^\prime)t} \langle \tilde{b}_i^\dagger(\omega) \tilde{b}_i(\omega^\prime) \rangle.
\end{equation}
From the quantum Langevin equation in frequency space, we find that the qubit field operator is related to the input field operator as
\begin{equation}
    \tilde{b}_i(\omega) =  \frac{\sqrt{\gamma_{\mathrm{R},i}}\tilde{b}_{\rm in}(\omega)}{i(\omega - \omega_0) + \gamma_{1,i}/2} \equiv \sqrt{\gamma_{\mathrm{R},i}} \chi (\omega) \tilde{b}_{\rm in}(\omega),
\end{equation}
where $\chi(\omega)$ is the susceptibility of the qubit. If we substitute this relation back into Eq.~\eqref{eq: integral P_e}, we obtain
\begin{equation}
    P_{e,i} = \iint \frac{d\omega}{2\pi}\frac{d\omega^\prime}{2\pi} e^{i (\omega - \omega^\prime)t} \gamma_{\mathrm{R},i} \chi^*(\omega)\chi(\omega^\prime) \langle \tilde{b}_{\rm in}^\dagger(\omega) \tilde{b}_{\rm in}(\omega^\prime) \rangle.
\end{equation}
For a stationary input field the spectral density $S_{\rm in}(\omega)$ is defined by 
\begin{equation}
    \langle \tilde{b}_{\rm in}^\dagger(\omega) \tilde{b}_{\rm in}(\omega^\prime) \rangle = 2\pi S_{\rm in}(\omega) \delta(\omega - \omega^\prime),
\end{equation}
where $\delta(\omega - \omega^\prime)$ is the Dirac delta function. Therefore, after integrating over $d\omega^\prime$, the expression for the excited state population simplifies to
\begin{equation}
    P_{e,i} = \int_{-\infty}^{+\infty} \frac{d\omega}{2\pi} \gamma_{\mathrm{R}, i} |\chi(\omega)|^2 S_{\rm in}(\omega).
\end{equation}
For a Markovian thermal flat noise spectrum, $S_{\rm in}(\omega)=N_i$ is constant, and we can integrate over the Lorentzian shape of the resonance to finally get
\begin{equation}
    P_{e,i} = \gamma_{\mathrm{R}, i} N_i \int_{-\infty}^{+\infty} \frac{1}{2\pi} \frac{d\omega}{(\omega - \omega_0)^2 + (\gamma_{1,i}/2)^2} = \xi_i N_i.
\end{equation}
Here, we introduced the factor $\xi_i = \frac{\gamma_{\mathrm{R}, i}}{\gamma_{1,i}}$. Inverting the equation, we can get the average photon number of the TMS mode $i$ from the qubit expectation value as
\begin{equation}
    N_i = \xi_i^{-1} P_{e,i} = \xi_i^{-1} \langle \sigma_i^+ \sigma_i^- \rangle.
\end{equation}
Equivalently, for the photon correlations between the modes $M$, we find the expression
\begin{equation}
    M = \frac{1}{4}\sqrt{\xi_1^{-1}\xi_2^{-1}} \big( \langle \sigma_1^{x}\sigma_2^{x}\rangle - \langle \sigma_1^{y}\sigma_2^{y}\rangle -i  \langle \sigma_1^{y}\sigma_2^{x}\rangle - i  \langle \sigma_1^{x}\sigma_2^{y}\rangle\big).
\end{equation}
Note that these expressions are only valid for a small qubit excitation probability, i.e., the regime in which we do not see effects of qubit saturation, and that in the case of a perfect bidirectional coupling, the prefactor is $\xi_i^{-1} = 2$.
\begin{figure}[tbp!]
    \centering
    \includegraphics{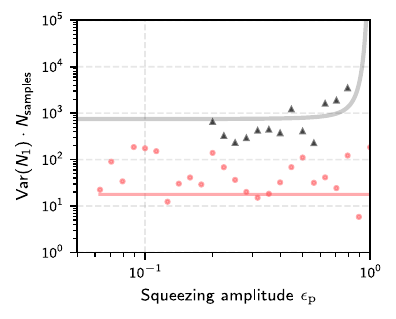}
    \caption{\textbf{Comparison of detection schemes.} Variance per shot for the estimate of $N_1$ using heterodyne detection (grey triangles) or the population of qubit 1 (red circles).}
    \label{fig:varofvar}
\end{figure}

To estimate the variances of $N_i$ and $M$, we remark that the qubit populations are inferred from a dispersive readout. We extract the expectation values from the resonator distributions corresponding to finding qubit $i$ in $\ket{g}$ or $\ket{e}$. Being $S_i^g$ ($S_i^e$) the average value for the resonator response in each of the cases and $\mathrm{Var} \{S_i\}$ the variance of the Gaussian distributions, we can define the SNR of the qubit readout as 
\begin{equation}
    \mathrm{SNR}_i = \frac{(S_i^g - S_i^e)^2}{\mathrm{Var} \{S_i\}},
\end{equation}
and the variance of the estimated photon number is then
\begin{equation}
    \mathrm{Var} \{N_i \} = \mathrm{SNR}_i^{-1} \xi_i^{-2}.
\end{equation}
In the case of qubit detection, the variance is independent of the photon number and only depends on the SNR of the dispersive readout. Fig.~\ref{fig:varofvar} shows the comparison of the two measurement protocols for the estimation of $N_1$ using qubit 1. We observe an improvement of 2 orders of magnitude in the single-shot variance, which explains why we can measure $N_1$ accurately for very low photon numbers, see Fig.~\ref{fig:4}. For qubit 2, we have not been able to fit the experimental variance by accounting only for the limited SNR of the dispersive readout, which points to an additional source of variance most likely coming from frequency instability.

\section{Limiting experimental factors}\label{app: limitations}
\begin{figure}[tbp!]
    \centering
    \includegraphics[width=\columnwidth]{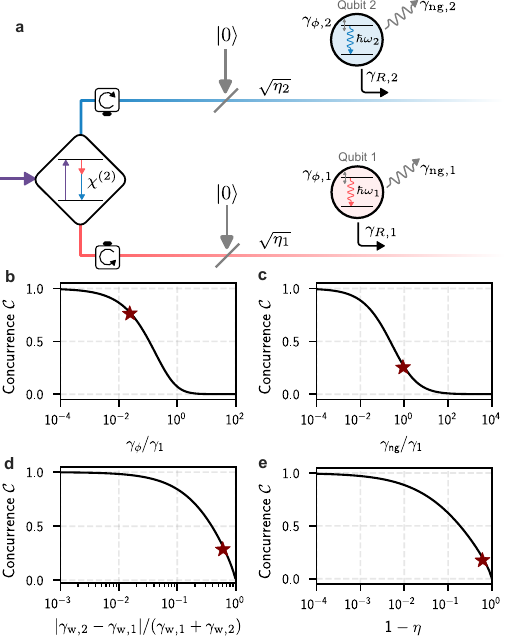}
    \caption{\textbf{Limiting experimental factors.} \textbf{a}, Realistic experimental setup indicating the relevant losses. Qubit-waveguide interaction is considered unidirectional. \textbf{b}, Concurrence as a function of qubit dephasing, \textbf{c}, non-guided losses, \textbf{d}, coupling asymmetry and \textbf{e.} photon loss. Stars indicate the corresponding parameters for this experiment.}
    \label{fig:limitingfactors}
\end{figure}

We use Eq.~\eqref{eq:App_Qubit_ME} to calculate the maximum available concurrence in the presence of experimental imperfections such as qubit dephasing, non-guided losses, asymmetry in the qubit-waveguide couplings and photon loss. A realistic experimental setup is depicted in Fig.~\ref{fig:limitingfactors}a, where we consider the case of chiral coupling to the waveguide. Experimentally, this scenario can be realized by using actual chiral interactions \cite{lodahl_chiral_2017, joshi_resonance_2023} or by coupling the qubits at the end of a terminated waveguide. For the simulations, we assume symmetric values for both qubits unless stated otherwise and optimize over $\epsilon_{\rm p} \in [0,1)$ to find the maximum concurrence for a given set of parameters.

It is possible to achieve a finite concurrence while the dephasing rate $\gamma_\phi$ is kept smaller than the qubit-waveguide coupling $\gamma_{\rm R} = \gamma_1 = \gamma_2$, maintaining $\gamma_{\rm ng} = 0$. For the values of Table~\ref{tab:qubit_params_booktabs}, we find that we could have achieved up to $\mathcal{C}^\star\approx 0.8$ if this were the only cause of imperfection. In principle, the ratio $\gamma_\phi/\gamma_1$ can be made smaller by increasing the coupling to the waveguide and reducing flux noise \cite{braumuller_characterizing_2020}.
In case of non-guided losses (or imperfect chiral interactions), we obtain the limit of $\mathcal{C}^\star \approx 0.25$ for the bidirectional waveguide case.
The qubit-waveguide coupling is mostly defined by the circuit geometry for a given transition frequency $\omega_{\rm q}$ and can be approximated by~\cite{houck_controlling_2008}
\begin{equation}
    \gamma_{\rm w} \approx \frac{\omega_{\rm q}^2 Z_0 C_c^2}{C},
\end{equation}
where $C_c$, $C$ are the coupling and total capacitance and $Z_0 = \SI{50}{\Omega}$ is the impedance of the line. In a planar architecture, careful design and simulation can arbitrarily reduce the asymmetry $|\gamma_{\mathrm{w},2} - \gamma_{\mathrm{w},1}|/(\gamma_{\mathrm{w},1} + \gamma_{\mathrm{w},2})$. 

Finally, we study the impact of photon loss. We can model the loss as $\eta_i = \eta_{\mathrm{JPC}, i}\cdot \eta_{\mathrm{path}, i}$ with two loss contributions arising from the JPC itself and from the path connecting the JPC and the qubits. The first one accounts for imperfections and internal losses in the generation of the TMS, which we find to be $\eta_\mathrm{JPC} \approx -(3, 8)$ dB, presumably due to device aging. The latter comes from the insertion loss of the hybrids, circulators, cables, and SMA and PCB connectors, and we estimate it  to be on the order of \SI{-1}{\dB} as suggested by other works~\cite{abdo_teleportation_2025, murch_reduction_2013}. With perfect squeezing generation and a moderate path loss of $1 - \eta = 0.1$, the maximum achievable concurrence is $\mathcal{C}^\star \approx 0.6$, comparable to existing stabilizing schemes~\cite{shah_stabilizing_2024-1}. 
Newer designs of JPCs show much higher squeezing and gain bandwidth~\cite{abdo_teleportation_2025-1} as well as other kinds of Josephson-based parametric amplifiers~\cite{eichler_quantum-limited_2014,qiu_broadband_2023} which, contrary to the JPC, have a single output port and do not generate spatially-separated entangled modes. It is also possible to generate a spatially-separated TMS state by interfering two single-mode squeezed states in a beam splitter~\cite{fedorov_finite-time_2018}.
Higher bandwidths would help to reach higher photon numbers and to extend the experiment to more qubits~\cite{agusti_autonomous_2023}, clearly leveraging the power of this protocol with respect to other autonomous protocols~\cite{shah_stabilizing_2024-1} for complex network architectures.

Additionally, we tackle the problem of preserving the entangled state. In the current configuration, the qubits decay into the waveguide on a timescale of approximately $T_1$ right after the stabilization drive has been turned off. In order to protect the state from decaying, different solutions are at hand: use tunable couplers to turn off the coupling between qubit and waveguide, swap the qubit state to a qubit not coupled directly to the transmission line~\cite{kannan_-demand_2023}, use qubits with additional ground states that are dark to the waveguide or couple more qubits in a chain via exchange interaction~\cite{zippilli_entanglement_2013, lingenfelter_exact_2024, irfan_loss_2024}.

\begin{figure}[t!]
    \centering
    \includegraphics[width = \columnwidth]{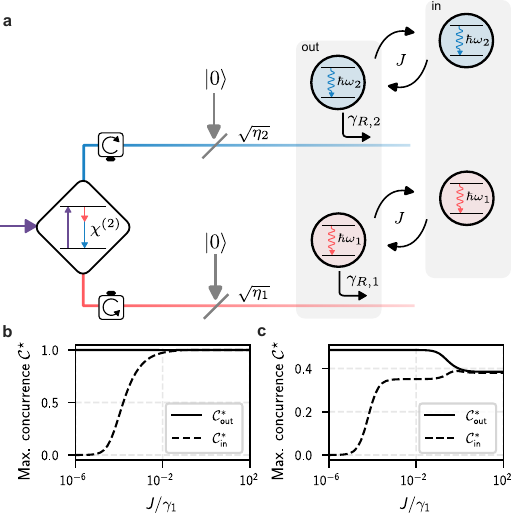}
    \caption{\textbf{Entanglement replication and storage.} \textbf{a}, Sketch of the extension to 4 qubits. The inner pair is not directly coupled to the waveguide. \textbf{b}, Concurrence of both pairs for different values of the exchange coupling $J$ for $\eta = 1$ and \textbf{c.} $\eta = 0.8$.}
    \label{fig:4qubits}
\end{figure}

We explore the latter approach and find that the inner qubits get entangled while the protocol remains autonomous. As shown in Fig. \ref{fig:4qubits}a, we extend the experiment considering two qubits on each site coupled via the exchange Hamiltonian
\begin{equation}
    H_J = \sum_{i = 1,2} J\big(\sigma_{i, \rm in}^+ \sigma_{i, \rm out}^- + \sigma_{i, \rm out}^+ \sigma_{i, \rm in}^-\big),
\end{equation}
while only the outer qubits are directly coupled to the waveguide. We find that the qubits are stabilized in a product state $\ket{\psi}\propto\ket{\Phi^+}_{\rm out} \otimes\ket{\Phi^-}_{\rm in}$ with $\ket{\Phi^\pm} = (\ket{gg} \pm \ket{ee})/\sqrt{2}$. In the case of perfect transmission $\eta = 1$ (Fig.~\ref{fig:4qubits}b), the concurrences of both pairs saturate to 1 as the exchange coupling approaches the qubit-waveguide coupling. However, for a finite transmission $\eta = 0.8$, the concurrence of the two qubit pairs is reduced with respect to the one achievable with a single qubit pair ($J/\gamma_1 \rightarrow 0$), in contrast to the protocols based on a coherent drive~\cite{lingenfelter_exact_2024, irfan_loss_2024}.

Thus, we predict that future implementations of this protocol with better squeezers and state-of-the-art fabrication can significantly increase the distributed concurrence. 
\newpage
\bibliography{TMSQlibrary_bib}
\end{document}